\documentstyle[12pt,aps]{revtex}
\tighten
\begin{document}
\title{{\Large {\bf 
Toward chiral theory of NN interactions in nuclear matter\bigskip}}}
\author{Boris Krippa\bigskip}
\address{ Department of Physics and Astronomy, Free University
of Amsterdam,\\ De Boelelaan 1081, 1081 HV Amsterdam.\\}
\maketitle
\vspace{1cm}
\begin{abstract}
 We consider an effective field theory of NN system
in nuclear medium. The shallow bound states, which complicate the 
effective field theory analysis 
in  vacuum 
do not exist in matter. We show that  
the next-to-leading order terms in the chiral expansion of the 
effective Lagrangian can  be interpreted as corrections so that
the expansion is systematic. The Low Energy Effective Constants
of this Lagrangian are found to satisfy the concept of naturalness. 
The potential energy per particle is
calculated. The problems and challenges in 
constructing the chiral theory of nuclear matter are outlined.

\end{abstract}

\vskip0.9cm


Effective Field Theory (EFT) is now a
standard method to study nuclear dynamics.
EFT is based on the use the Lagrangian
 with the appropriate effective degrees of freedom instead
of the fundamental ones in the low-energy region (for review of EFT see,
for example \cite {Ma95}). This Lagrangian  includes all possible terms 
allowed by the symmetries of the underlying theory.  The states which
 can  be treated as heavy, compared to the typical energy scale involved,
are integrated out. They are hidden in the Low Energy Effective
Constants (LEC's) of Lagrangian.
 The physical amplitudes can be represented as the sum 
of certain graphs, each of them being of a given order in $Q/\Lambda$,
where $Q$ is a typical momentum scale and $\Lambda$ is a scale 
 of the short range physics.
 The relative contribution of each graph can  roughly be estimated
using  counting rules \cite {We79}. However, being applied to the
 NN system EFT encounters
serious problem which is due to existence of the  bound
states near threshold \cite {We91}. It results in the large nucleon-nucleon
scattering length and makes the perturbative expansion divergent. Weinberg 
suggested \cite {We91} to apply  counting rules to the certain class 
of the irreducible diagrams which should  then be summed up to infinite 
order by solving the Lippmann-Schwinger (LS) equation. 
The irreducible diagrams can be treated as the effective potential
 in this case. Different aspects of the chiral NN problem 
have been discussed since then \cite {Ka}. The  EFT method
has also  been used to study  nuclear matter 
\cite {Se97,Ly,Lu,Fr}. In \cite {Se97} the effective
 chiral Lagrangian was constructed and the ``naturalness'' of the
 effective coupling constants has been demonstrated. The possible
counting rules for nuclear matter have been discussed in \cite {Lu}.
These two lines of development of the chiral nuclear physics are in some sense
similar to the tendencies existed some time ago in  conventional nuclear
physics with the phenomenological NN forces. On the one hand,
 the phenomenological NN potentials were used to describe
 nucleon-nucleon cross sections
and phase shifts. On the other hand,  nuclear mean field approaches
 provided a reasonable description of the bulk
 properties of nuclear matter.
 The unification of these two approaches then led to the famous
Bethe-Goldstone (BG) equation \cite {Be} for the G-matrix which is an 
analog of  scattering T-matrix, satisfying the LS equation.  
So one may follow the same strategy and, being
 equipped with
the chiral theory of NN interaction in vacuum, try to construct the chiral
G-matrix.
One can easily see the qualitative difference between vacuum and medium cases.
 In nuclear medium because of Pauli blocking  the intermediate 
states with the momenta
less than Fermi momentum $p_F$ are forbidden. Therefore, the nucleon propagator
does not exhibit a pole. Moreover, the shallow bound  NN states,
being a serious problem in vacuum, simply do not
exist in nuclear matter because of  interaction of the NN pair with
 nuclear mean
 field. It means that the effective scattering length becomes considerably
smaller compared to the vacuum one. 
. The moderate value of the in-medium scattering length would
 indicate  that the typical scale 
 of the NN interactions 
gets ``more natural'' in nuclear matter.      
We start from the standard nucleon-nucleon effective chiral Lagrangian
\begin{equation}
{\cal L}=N^\dagger i \partial_t N - N^\dagger \frac{\nabla^2}{2 M} N
- \frac{1}{2} C_0 (N^\dagger N)^2\\ 
-\frac{1}{2} C_2 (N^\dagger \nabla^2 N) (N^\dagger N) + h.c. + \ldots.
\label{eq:lag}
\end{equation}
We consider the NN scattering in the $^{1}S_0$ state.
The G-matrix is 
\begin{equation}
G(p',p)=V(p',p) + M \int \frac{dq q^2}{2 \pi^2} \, V(p',q) 
\frac{\theta(q-p_F)}{M(\epsilon_{1}(p) +\epsilon_{2}(p')) - q^2} G(q,p),
\label{eq:LSE2}
\end{equation}
Here $\epsilon_1$ and $\epsilon_2$ are the single-particle 
energies of the bound nucleons.
For the in-medium nucleon mass we used the value $M = 0.8 M_0$, where
 $M_0$ is the nucleon mass in vacuum. 
 The standard strategy of  treating the chiral NN problem in vacuum
is the following. One computes amplitudes up to a given chiral order
in the terms of the effective constants $C_0$ and $C_2$ which are then 
determined by comparing the calculated amplitude with some experimental data.
 Having these constants fixed one
can calculate the other observables. 
We will follow the similar strategy  in the nuclear matter case  
and proceed as follows.
We take exactly solvable separable potential with parameters adjusted
to the value of the potential energy per particle in nuclear matter
and will consider the G-matrix obtained from this separable potential
as our ``observable''.
 Then we  solve the BG equation
with the effective constants $C_0$ and $C_2$. The numerical values 
of these constants are determined comparing the phenomenological
and EFT G-matrix at some fixed kinematical points.
The check of consistency we used is the 
 difference between $C_0$'s
determined in the leading and subleading orders. 
In the vacuum case
the this difference was found to be large \cite{Co97}.
Using a separable potential with the effective strength $\lambda$
and the form factors
\begin{equation}
\eta(p)={(p^2 + \beta^2)^{-1/2}}
\end{equation}
 One can easily find the solution of the corresponding BG equation
\begin{equation}
G(k,k)=-
\eta^{2}(k)\left[\lambda^{-1}+ 
\frac{M}{2 \pi^{2}} \int{dqq^2} \, \frac{\theta(q-p_F) \eta^{2}(q)}{{k^2}- q^2}\right]^{-1}
\end{equation}
We choose 
$\lambda = 2.4$ and $\beta = 1.1\, {\rm fm}$
to  provide 
the potential energy per particle in a good agreement with the empirical value. The parameters
 $ \lambda$ and $\beta$ being substituted in the G-matrix lead  to  
$a_m\simeq r_m\simeq 0(1)$, where $a_m$ and
  $r_m$ are the in-medium analogs of scattering length and effective radius.
 The absolute value of the  in-medium scattering length is considerably reduced 
compared to the vacuum one. It clearly indicates that, as expected, the 
shallow virtual nucleon-nucleon bound state is no longer present in nuclear medium.
Thus,  one can avoid significant part of the difficulties typical for the chiral NN
problem in vacuum. Having determined the phenomenological G-matrix
one can now solve the BG equation taking into account leading and sub-leading
orders of the NN effective chiral Lagrangian. The solution is similar to 
the vacuum case \cite{Co97}   
\begin{equation}
\frac{1}{G(k,p_F)}=\frac{(C_2 I_3(k,p_F) -1)^2}{C_0 + C_2^2 I_5(k,p_F)
 + {k^2} C_2 (2 - C_2 I_3(k,p_F))} - I(k,p_F),
\label{eq:Tonexp}
\end{equation}
where we defined
\begin{equation}
I_n \equiv -\frac{M}{2 \pi^2} \int dq q^{n-1}\theta(q-p_F);
I(k) \equiv \frac{M}{2 \pi^{2}}\int dq  \, \frac{q^2\theta(q-p_F)}
{{k^2}- {q^2}}.\label{eq:IEdef}
\end{equation}
These integrals are divergent so the renormilization should be carried out.
The procedure used is similar to that adopted in Ref. \cite {Ge} to study the 
EFT approach to the NN interaction in vacuum. We subtract the divergent integrals at
some kinematical point $p^2 = -\mu^2$. After subtraction the renormalized G-matrix takes the form
\begin{equation}
\frac{1}{G^{r}(k,k)}=\frac{1}{C^{r}_{0}(\mu) 
 +  2{k^2} C^{r}_{2}(\mu)} +\frac{M}{4\pi}[ p\log\frac{p_F - p}{p_F + p}
 - i\mu\log\frac{p_F - i\mu}{p_F + i\mu}] ,
\label{eq:GR}
\end{equation}
One notes that in the $p_F \rightarrow 0$ limit  the vacuum chiral NN amplitude is recovered.
We choose the value $\mu$ = 0 as a subtraction point. The $\mu$ dependence of LEC's is 
governed by the renormalization group (RG) equations. We demand that the entire G-matrix is
independent of substruction point. After differentiating the G-matrix with respect to 
$\mu$ and setting $\partial G/\partial \mu$ = 0 one can get the following RG equations
\begin{equation}
\frac{\partial C_0(\mu)}{\partial \mu} = \frac{C_0^2 M}{4\pi^2}(\frac{\mu p_F}{p_F^2 + \mu^2}
+ 2 tan^{-1}(\frac{\mu}{p_F}))
\end{equation} 
\begin{equation}
\frac{\partial C_2(\mu)}{\partial \mu} = \frac{C_0 C^2 M}{\pi^2}(\frac{\mu p_F}{p_F^2 + \mu^2}
+  tan^{-1}(\frac{\mu}{p_F}))
\end{equation} 
In the limit $p_F \rightarrow$ 0 these equations transform to the ones derived 
by Kaplan et al. \cite{Ka}. 
 Now one can determine the LEC's by equating
the EFT and phenomenological G-matrices at some kinematical points. We used the values 
$p = \frac{p_F}{2} ; \frac{p_F}{3}$ as such points. The assumed value of the Fermi-momentum
is $p_F$ = 1.37 fm. In the following we will omit the label ``r'' implying that we always
 deal with 
renormalized quantities.  
We found $C_0 = -1.86 fm^2$ in LO. In NLO one gets $C_0= -2.7 fm^2$ and $C_2 = 0.84 fm^4$
so that the inclusion of the NLO corrections  give rise to the approximately 40$\%$ change
in the value of  $C_0$. It indicates that the chiral expansion is systematic in a sense that
 adding of the NLO 
terms in the effective Lagrangian brings in a ``NLO change'' of the coefficients which have 
already 
been fixed at LO. The natural size of the in-medium scattering length and moderate changes
experienced by the coupling constant $C_0$ might, in principle, indicate the possibility
of the perturbative calculations.   However, in spite of
this, it is still more
useful to treat this problem in the nonperturbative manner since the corrections themselves
are quite significant. Moreover, the overall (although distant)
goal of the EFT description is to derive both nuclear matter and the 
vacuum NN amplitude from the same Lagrangian. However, it is hard to say 
at what densities the dynamics becomes intrinsically nonperturbative, so it is 
better to treat the problem nonperturbatively from the beginning. 
Let's now calculate the potential energy per particle
 using the expression 
 \begin{equation}
U_{tot} = \frac{1}{2}\sum_{\mu,\nu}<\mu\nu|G(\epsilon_\mu + \epsilon_\nu)
|\mu\nu-\nu\mu)
\end{equation}
The summation goes over the states with momenta below $p_F$.
 The LO perturbative 
calculations (where $G \simeq C_0$) give  $\frac{U(^1S_0)}{A}\simeq -12 MeV$
The calculations using the lowest 
order $G$-matrix result in the value $\frac{U(^1S_0)}{A}\simeq -17 MeV$.
The inclusion of the next-to-leading order corrections gives rise to the 
value $\frac{U(^1S_0)}{A}\simeq -13.1 MeV$. 
Similar calculations done in the triplet s-wave channel give rise  to the value
$\frac{U(^3S_1)}{A}\simeq -17.3 (-13.2)$ MeV in LO (NLO). The values of the potential
 energy obtained with 
chiral approach looks quite reasonable although they are somewhat smaller 
than the standard values usually obtained in the calculations with the 
phenomenological two-body forces \cite {Tab}. Thus,  one can  conclude that
there is still a room for both pionic effects and many body  correlations.
 One notes, however, that adding pions 
will average to NLO effects. It agrees with the results of Refs. \cite{Fu97,Ru97,SeW} where
 a good fit of nuclear properties was obtained in the 
framework of the effective Lagrangians with the point-like interactions. Many body
forces are also expected to result in  rather small corrections. It follows both
from the   nuclear phenomenology and from Weinberg counting rules \cite{We91}.
So it is reasonable to expect that the inclusion of pion effects and many-body interactions
could change the exact value of the LEC's by some factor of order unity keeping their order 
of magnitude the same. It is interesting to see whether the values of LEC's obtained above
satisfies the naturalness criteria elaborated for the nuclear matter case in \cite{Se97}.
According to the concept of naturalness as formulated in \cite{Se97} an individual term in
the effective Lagrangians can schematically be written as
\begin{equation}
c\left[\frac{\psi^{+}\psi}{f_{\pi}^2\Lambda}\right]^l\left[\frac{\partial}{\Lambda}\right]^n
 (f_\pi\Lambda)^2.
\end{equation}  
Applying the scaling rules developed in \cite{Se97} to extract
 all dimensionful factors and assuming that $\Lambda \simeq$ 600 MeV
one finds $c_0(c_2) \simeq 0.7 (0.65)$. Thus the 
dimensionless coefficients are indeed compatible with naturalness. As was pointed out in
\cite{Se97} this is a nontrivial fact as the terms in the effective Lagrangian are
 supposed to absorb long distance effects from the ladder and ring diagrams.

The validity of the EFT description is restricted by some cutoff parameter $\Lambda$,
following from naturalness and 
reflecting the short range physics effects. Its value deserves some comments 
in the context of applying of the  EFT methods to nuclear matter. The scale 
where the EFT treatment ceases to be valid should approximately correspond 
to the scale of the short range correlations (SRC),
 that is, $\sim 500 - 600 MeV$.
The description of SRC is hardly possible in the framework of EFT so the value 
$\Lambda \sim 500 - 600 MeV$ might put natural constraint on the EFT description 
of nuclear matter. To make the chiral expansion meaningful the chiral counting rules 
in nuclear matter must be established. This is still open problem. However, 
the above obtained results suggest that the relevant expansion parameter 
could be something like $\frac{<p>}{\Lambda} \sim \frac{m_\pi}{\Lambda} \sim 0.3 -0.4$,
 where $<p>$ is the nucleon average momentum in nuclear matter.
 Of course, until pion effects and many-body 
interactions are taken into account this estimate can only be suggestive.

\end{document}